# Automated spacing measurement of formwork system members with 3D point cloud data

Keyi Wu [a], Samuel A. Prieto [a], Eyob Mengiste [a], and Borja García de Soto [a]

[a] S.M.A.R.T. Construction Research Group, Division of Engineering, New York University Abu Dhabi (NYUAD), Experimental Research Building, Saadiyat Island, P.O. Box 129188, Abu Dhabi, United Arab Emirates.
E-mail: keyi.wu@nyu.edu, samuel.prieto@nyu.edu, eyob.mengiste@nyu.edu, garcia.de.soto@nyu.edu

**Abstract**

The formwork system belonging to the temporary structure plays an important role in the smooth progress and successful completion of a construction project. Ensuring that the formwork system is installed as designed is essential for construction safety and quality. The current way to measure the spacing between formwork system members is mostly done using manual measuring tools. This research proposes a framework to measure the spacing of formwork system members using 3D point cloud data to enhance the automation of this quality inspection. The novelty is not only in the integration of the different techniques used but in the detection and measurement of key members in the formwork system without human intervention. The proposed framework was tested on a real construction site. Five cases were investigated to compare the 3D point cloud data approach to the manual approach with traditional measuring tools. The results indicate that the 3D point cloud data approach is a promising solution and can potentially be an effective alternative to the manual approach.

## 1. Introduction

In a construction project, the formwork system is a set of temporary structures used to mold fresh concrete into the desired appearance, shape, dimension, and location [1]. As a critical element in concrete construction, the formwork system can not only easily cause safety risks during concrete placement (in the case of collapse due to improper installation) but also seriously affect the quality of resulting concrete structures after the placement [2]. The formwork system occupies a significant proportion of the concrete construction investment, and it is reported to usually account for 40% to 60% of the cost of concrete construction [3]. Overall, the formwork system plays an important role in the smooth progress and successful completion of a construction project [4].

Many aspects of the formwork system, such as loads, methods, and materials, are different from those of permanent structures [5]. Compared to permanent structures, the formwork system is more fragile and requires sufficient strength, stiffness, and stability [6]. The quality inspection of the formwork system is an essential construction management procedure that examines, measures, and gauges whether the characteristics (e.g., number, size, spacing) of the installed formwork system properly comply with specific design requirements [5]. If the formwork system does not satisfy design requirements, it may not be strong enough and thus cause severe damage or deficiencies to concrete structures, even fatalities in the case of a collapse; and if the formwork system exceeds design requirements, it may result in unnecessary waste, such as more resources and cost [4]. Therefore, the quality inspection of the formwork system is of great practical significance in construction projects [5].

The formwork system typically comprises different categories of members, including panels, studs, wales, ties, and braces, which are highlighted in yellow, purple, green, red, and blue, respectively, in Figure 1. To ensure the strength, stiffness, and stability of the formwork system, the spacing of studs, wales, ties, and braces needs to be specifically designed according to the force determined by the properties of concrete (e.g., density, initial setting time, slump) and the means and methods of construction (e.g., placement approach, speed), as well as the object determined by the features of members (e.g., material, size, shape) [4][5]. Generally, the traditional spacing measurement of formwork system members is mainly conducted using manual measuring tools (e.g., measuring tape, laser distance meter). This process is time-consuming and labor-demanding; in addition, it is easily affected by the inspector's knowledge and skills, making the collection of data subjective and inconsistent [4][7].

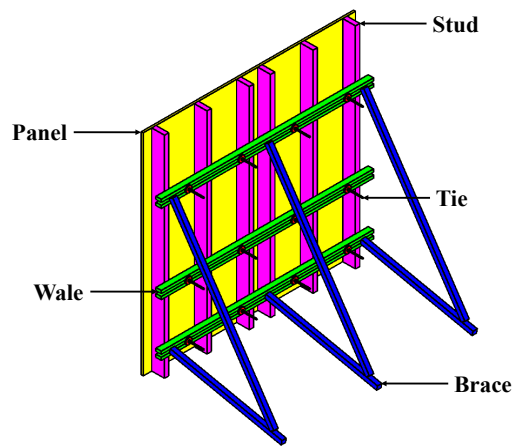

*Figure 1. A typical formwork system and related categories of members.*

To overcome the above shortcomings of the traditional spacing measurement, other approaches should be investigated. A 3D point cloud is a set of data points containing three-dimensional Cartesian coordinates (X, Y, Z) and other information such as color (R, G, B) and reflectance or intensity [8][9]. In this sense, construction site scenes can be rapidly reconstructed in terms of the 3D point cloud. Specifically, the external surface of the formwork system can be effectively represented by data points in the 3D point cloud, and it is possible to automate the detection and measurement of different members in the formwork system using the three-dimensional Cartesian coordinates (X, Y, Z) contained in the data points. For this reason, this research proposes a framework for the spacing measurement of formwork system members with 3D point cloud data. The rest of this article is structured as follows. Section 2 reviews the literature on the acquisition approach of 3D point cloud data and its application in dimensional quality inspection in construction projects. Section 3 elaborates on the two parts with eight steps of the proposed framework, and a case study is conducted for its demonstration in Section 4. Section 5 explores the significance of the proposed framework. In Section 6, the conclusion and outlook are provided.

## 2. Literature review

### 2.1 Acquisition approach of 3D point cloud data

3D point cloud data can be acquired through various approaches, including laser scanning,

photogrammetry, and videogrammetry [12]. Laser scanning is an approach to record object information according to dense points detected by reflected laser beams [13][14]. In laser scanning, Time-of-Flight (ToF) and Phase-Shift (PS) are the main techniques for range measurement [15]. ToF uses the time interval between emitting and receiving a laser beam and the velocity of the laser to estimate the distance. PS applies the phase difference between emitting and receiving a laser beam and the wavelength of the laser to infer the distance. Photogrammetry is an approach to record object information according to photographic images and patterns of radiant electromagnetic energy and other phenomena [16][17]. In photogrammetry, Structure from Motion (SfM) and Multi-View Stereo (MVS) are the primary methods for point cloud generation [15][18]. SfM matches feature points extracted from images to perform a sparse three-dimensional reconstruction. MVS corrects images according to camera parameters output from SfM and matches feature points extracted from the images to perform a dense three-dimensional reconstruction. Videogrammetry is an approach similar to photogrammetry but records object information from video streams [12]. In videogrammetry, the three-dimensional reconstruction is determined by video images taken from different angles [15].

Generally, laser scanning has a higher measurement accuracy, a longer measurement range, less chance for user errors, and is not affected by lighting, whereas photogrammetry and videogrammetry have a lower cost and better visual representation of textures [19]. Since the installation quality of the formwork system is significantly related to the spacing of members, there is a relatively high requirement for the accuracy of the spacing measurement in construction projects. Meanwhile, the dynamic and complex nature of construction sites increases the uncertainty in the spacing measurement, requiring the data acquisition process to be more flexible and robust. In addition, the installation of the formwork system typically takes place outdoors; therefore, the data acquisition process is extremely susceptible to the influence of lighting. Keeping these points in mind, laser scanning is used for this research because it is more in line with the requirements and characteristics of the spacing measurement of formwork system members.

**2.2 Dimensional quality inspection with 3D point cloud data**
3D point cloud data has been effectively applied to various dimensional quality inspections in construction projects. Some of the most relevant applications are summarized in terms of field, object, and inspection item in Table 1.

Precast components have become one of the most concerned fields with increasing demands on dimensional quality control during the fabrication stage, mainly involving objects such as reinforcing bars, concrete slabs and panels, and bridge deck slabs and girders. For reinforced precast components, reinforcing bars at connections significantly impact overall structural integrity. To guarantee that reinforcing bars are installed at the intended positions, a colored point cloud-based approach was provided to examine reinforcing bar positions [20]. Concrete slabs and panels are commonly used precast components, and they would cause structural failure if dimensions are not properly controlled. For assessing the dimensional quality of concrete slabs, a method was proposed to estimate slab sizes and shear pocket sizes and positions [21]. Since the openings of precast components increase the difficulty of dimensional quality control, a technique was presented to detect panel sizes and squareness and hole sizes and positions for accelerating the inspection of concrete panels with openings [22]. Precast bridge deck slabs and girders are widely used in bridge construction, and they are connected through shear pockets on deck slabs and connectors on girders. To ensure the proper connection between bridge deck slabs and girders, a method was provided to check slab sizes, girder sizes, shear pocket sizes and positions, and shear connector orientations and positions [23]. Considering that irregular precast components can increase the complexity of dimensional quality inspection, an approach was developed to assess the panel depth, shear key sizes and positions, and flat duct positions of a bridge deck slab with geometric irregularities [24]. The space for dimensional quality inspection is usually limited due to the close placement of precast components. Therefore, a mirror-aided methodology was proposed to estimate the outer sizes and shear key sizes of the side surface of a bridge deck slab [25].

*Table 1. Application of 3D point cloud data for dimensional quality inspection in construction projects.*

| Source | Field | Object | Inspection item |
| --- | --- | --- | --- |
| [20] | | Reinforcing bars | Reinforcing bar positions |
| [21] | | Concrete slabs | Slab sizes; Shear pocket sizes and positions |
| [22] | | Concrete panels | Panel sizes and squareness; Hole sizes and positions |
| [23] | Precast components | Bridge deck slabs and girders | Slab sizes; Girder sizes; Shear pocket sizes and positions; Shear connector orientations and positions |
| [24] | | Bridge deck slabs | Panel depth; Shear key sizes and positions; Flat duct positions |
| [25] | | Bridge deck slabs | Outer sizes; Shear key sizes |
| [26] | Reinforced concrete components | Reinforcing bars and formwork | Reinforcing bar spacing; Concrete covers; Formwork inner sizes |
| [27] | | Reinforcing bars | Reinforcing bar sizes and spacing |
| [28] | | Reinforcing bars | Reinforcing bar spacing |
| [29] | Steel structure components | Steel columns | Column positions and altitudes |
| [30] | Spatial structure components | Free form structure elements | Element sizes and angles |
| [31] | Temporary structure components | Scaffolding platforms | Toe-board heights; Guard-rail positions |
| [32] | | Formwork supporting frames | Pole spacing; Tube spacing |

Other components have also received attention, including those for reinforced concrete, steel structures, spatial structures, and temporary structures. The quality of reinforced concrete components depends largely on the correct installation of reinforcing bars and formwork. To monitor the placement of reinforcing bars in formwork, a method was designed to detect reinforcing bar spacing, concrete covers, and formwork inner sizes [26]. Since the bearing capacity of reinforced concrete components depends on the diameter and position of reinforcing bars, a methodology was proposed to classify reinforcing bar sizes and estimate reinforcing bar spacing [27], and an approach was presented to assess reinforcing bar spacing [28]. Taking into account the inadequacy of as-built dimension tracking solutions, a technique was proposed to examine the bottom and top center point positions and altitudes of steel columns for dimensional compliance control [29]. To enhance the quality control accuracy and efficiency of complex spatial structure components, a method was developed to detect the sizes and angles of free-form structure elements [30]. For fall prevention systems used to ensure the safety of a scaffolding platform, a methodology was developed to evaluate the toe-board heights and guardrail positions to meet scaffolding safety regulations [31]. To reduce safety risks related to the improper installation of formwork supporting frames, an approach was presented to estimate pole spacing and tube spacing to comply with design requirements [32].

Overall, many dimensional quality inspections in construction projects have been beneficially explored with 3D point cloud data. However, the spacing measurement of formwork system members has received little attention. Different from the objects mentioned above that have been studied (e.g., reinforcing bar, concrete slab, steel column, scaffolding platform), since there are specific force transfer features, material properties, and design standards for the formwork system, it has unique members and layouts. A typical formwork system is composed of panels, studs, wales, ties, and braces (Figure 1) that are installed following a fixed relative position order, which makes it difficult for previous studies to be applied to such members and layouts. Therefore, a specific methodology for the spacing measurement of formwork system members using 3D point cloud data needs to be developed. In particular, considering the purpose of enhancing the automation of the spacing measurement of formwork system members, it is essential and meaningful to conduct a specialized study for the segmentation, identification, and recognition of members according to the characteristics of the formwork system.

## 3. Proposed framework

This section elaborates on the proposed framework for the spacing measurement of formwork system members with 3D point cloud data. This research focuses on the preprocessing, processing, and analysis of 3D point cloud data; therefore, the data acquisition process is out of the scope (i.e., 3D point cloud data has already been obtained). The proposed framework is divided into two parts: (1) Part 1: 3D point cloud data preprocessing and (2) Part 2: 3D point cloud data processing and analysis (Figure 2). Each part is detailed below.

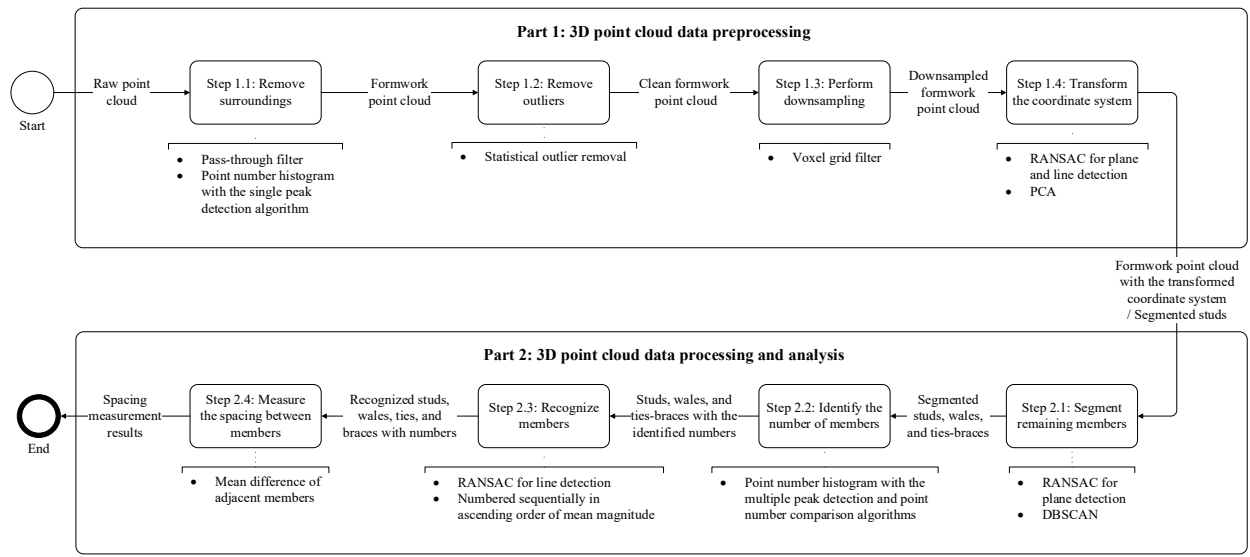

Figure 2. Proposed framework for the spacing measurement of formwork system members with 3D point cloud data.

## 3.1 Part 1: 3D point cloud data preprocessing

During the 3D point cloud data preprocessing, there are four steps: (1) Step 1.1: Remove surroundings, (2) Step 1.2: Remove outliers, (3) Step 1.3: Perform downsampling, and (4) Step 1.4: Transform the coordinate system, to be executed.

The first step is to remove the surroundings around the formwork system to avoid interfering with the spacing measurement. The point cloud acquired by laser scanning typically contains not only information on formwork system members but also on all surroundings due to the larger range of laser scanners (over 100 meters in most cases). The surroundings are of no use for the spacing measurement and should be removed, including the formwork panels that are not related to the spacing measurement, even though they are part of the formwork system. Since the number of points that belong to the surroundings is generally substantial, a pass-through filter that enables fast filtering is applied to remove them. Through conforming the 3D coordinates of the studied formwork system within the point cloud, the desired points inside or outside the set threshold of 3D coordinates are retained, and the other points are removed by the pass-through filter. The coordinates that act as thresholds for the pass-through filter are determined by manually extracting the corners of the studied formwork system in any point-cloud viewing software like CloudCompare [33]. In some cases, the ground is close to formwork system members, which makes it difficult to be removed completely. Generally, the number of points that belongs to the ground is significantly higher than that of formwork system members along the normal direction of the ground. In other words, a histogram of the number of points along the Z-axis of the point cloud will have different peaks. The highest peak in the histogram would represent the ground level (i.e., a higher number of points), and it can be easily distinguished from lower peaks representing the formwork system members close to the ground (but with a lower number of points). For this reason, the single peak detection algorithm can be applied to remove the ground. To generalize this and consider the cases when other elements are close to the ground (e.g., clutter or other materials on the ground next to the formwork system), the size of bins can be modified to include the points of these elements depending on the type of conditions investigated. For example, when the size of the bins is set to one unit, the second-highest peak representing the undesired elements would not be removed by the single peak detection algorithm (Figure 3 (a)). When the size of the bins is adjusted to two units, those undesired elements are placed at the highest peak (Figure 3(b)); thus, they can be removed successfully.

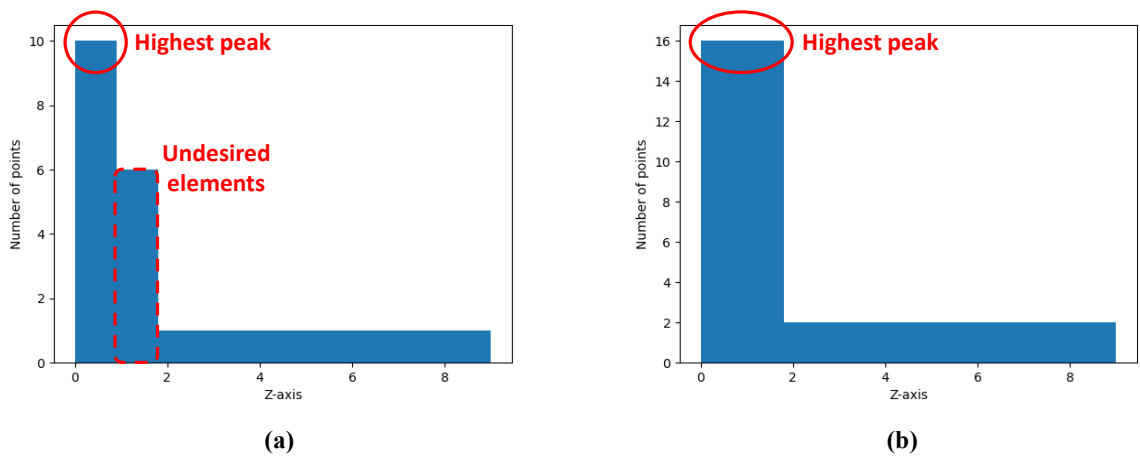

Figure 3: Modification of the size of bins for removing undesired points close to the ground (the size of the bins is one unit (a) and two units (b))

The second step is to remove the point cloud outliers that can adversely affect the spacing measurement of formwork system members. Laser scanning often produces point cloud outliers due to the presence of highly reflective surfaces, see-through objects, or around sharp corners where laser beams do not reflect properly. Although the number of point cloud outliers is not large compared to the total number of points, this complicates data processing and analysis as well as generates erroneous values, which may lead to the failure of the spacing measurement of formwork system members, on top of increasing computational time and resources processing points that do not add any semantic value. Considering that point cloud outliers are featured by being sparsely distributed in space, the statistical outlier removal approach [34], based on the computation of the distribution of distances between points and their neighbors, is employed to remove point cloud outliers. For each point, the mean distance to the set number of nearest neighbors is computed. The resulting mean distances of all points are assumed to follow a Gaussian distribution. If the mean distance of a point is outside the interval defined by the mean and the standard deviation of the Gaussian distribution, it is considered an outlier and removed from the point cloud.

The third step is to perform the downsampling of the point cloud to further reduce computational effort and errors. Laser scanning may produce overly dense and unevenly distributed point clouds, which can burden data processing and mislead data analysis, and is not conducive to the spacing measurement of formwork system members. To mitigate such negative effects, a voxel grid filter that enables downsampling is utilized to decrease the size of the point cloud and make the distribution of the point clouds more uniform. By setting a voxel size, points that fall within the cubic voxel are replaced with a centroid point, approximately representing the replaced points, to reduce the density of the point cloud without losing important features. It should be noted that the reason for performing downsampling after removing outliers is that the points of some small-sized formwork system members (e.g., ties) may be removed if the order is reversed.

The fourth step is to transform the coordinate system of formwork system members so that it can facilitate the spacing measurement. The coordinate system of the point cloud acquired by laser scanning usually does not adapt well to the orientation of formwork system members, which complicates the process of segmentation, identification, and recognition. Since formwork system members always follow a set of specific orientations, the process can be simplified if the coordinate system of the point cloud is aligned with those orientations. For the transformation of the coordinate system, the front surface of a stud with a rectangular shape feature is a suitable reference object because its long sides are significantly longer than its short sides, and thus the orientation of the transformed coordinate system is easy to identify. Considering that the plane corresponding to the front surface of all studs is the one with the maximum number of points, the random sample consensus algorithm (RANSAC) for plane detection is applied to detect the front surface of all studs first (Figure 4 (a)). Then, the RANSAC for line detection is applied to detect the front surface of a single stud in the detected front surface of all studs (Figure 4 (b)). Regarding the RANSAC for plane or line detection [35], in each iteration, points that can fit a plane or a line mathematical model are randomly selected first; then, the selected points are used to compute the parameters of the plane or line mathematical model; and finally, all points that fit the resulting plane or line model within the set threshold are detected as the front surface of all studs or the front surface of a single stud. In order to determine the transformed coordinate system, the principal component analysis (PCA) is applied according to the detected stud front surface (Figure 4 (c)). The first, second, and third principal axes of the transformed coordinate system correspond to the long side, short side, and normal directions of the detected stud front surface, respectively. With respect to the PCA [36], the first principal axis of the transformed coordinate system is determined so that the distribution variance of the projection of all the points of the detected stud front surface is maximized on this coordinate axis (i.e., the first principal component). The second principal axis is determined so that the distribution variance is maximized on the one that is perpendicular to the first coordinate axis (i.e., the second principal component). Finally, the last principal axis is the one perpendicular to the first and second principal axes (i.e., the third principal component).

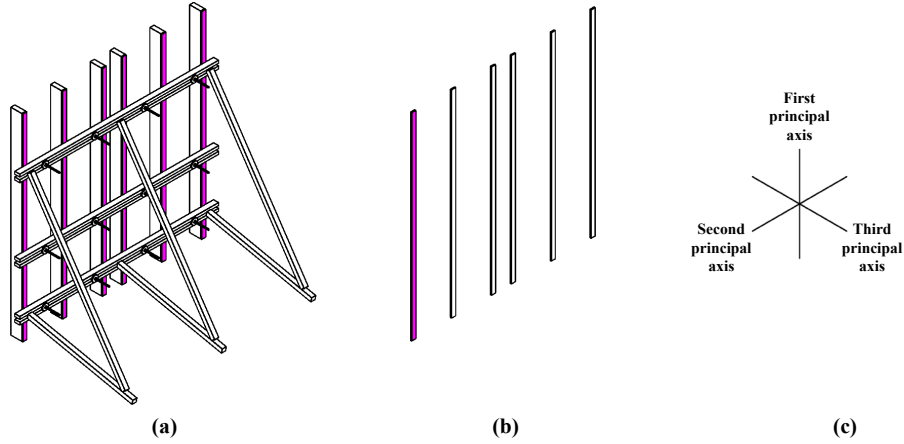

*Figure 4. Transformation of the coordinate system of formwork system members: (a) detection of the front surface of all studs (highlighted in purple) applying the RANSAC for plane detection; (b) detection of the front surface of a single stud (highlighted in purple) applying the RANSAC for line detection; and (c) determination of the transformed coordinate system applying the PCA.*

**3.2 Part 2: 3D point cloud data processing and analysis**
During the 3D point cloud data processing and analysis, there are four steps: (1) Step 2.1: Segment remaining members, (2) Step 2.2: Identify the number of members, (3) Step 2.3: Recognize members, and (4) Step 2.4: Measure the spacing between members, to be executed.

The first step is to segment the remaining categories of formwork system members (i.e., wales, ties, and braces) based on the transformed coordinate system. Various categories of formwork system members follow a well-defined force transfer path. The force that panels bear is transferred to studs, wales, ties and braces, which are installed following a fixed relative position order along the third principal axis. Since studs have been segmented before transforming the coordinate system, only wales and ties-braces need to be segmented in this step. For wales that do not overlap with ties and braces along the third principal axis, considering that the plane corresponding to the front surface of all wales is the one with the maximum number of points after removing the front surface of studs, the RANSAC for plane detection is employed for its segmentation (Figure 5 (a)). For ties and braces that overlap with each other along the third principal axis, they are segmented employing the density-based spatial clustering of applications with noise (DBSCAN) [37] (Figure 5 (b)). Regarding the DBSCAN, in each cluster, points that meet a set minimum number of points within the set distance are identified as core points of a tie or brace first; then, points that are connected to the core points are identified as border points of the tie or brace; and finally, points that are not connected to any core points are identified as outliers of the tie or brace. To facilitate the segmentation of ties and braces, the direction of the third principal required for removing the points belonging to other formwork system members axis (i.e., whether the points lie on the positive or the negative direction of the axis), needs to be identified. Since formwork system members are installed in turn along the same direction of the third principal axis, the mean of the coordinates of the points of the detected front surfaces for the stud and wale on the third principal axis are compared and used to identify the axis direction (Figure 5 (c)). For example, if the mean of the coordinates of points of the detected front surface for the stud (highlighted in purple) is smaller than that of the detected front surface for the wale (highlighted in green), it implies that formwork system members are installed along the positive direction of the third principal axis; otherwise, they are along the negative direction of the third principal axis.

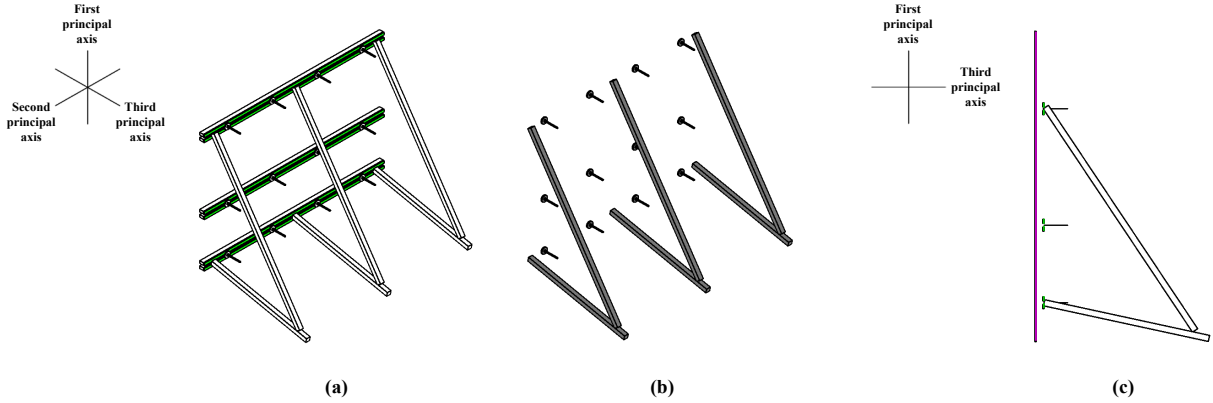

*Figure 5. Segmentation of different categories of formwork system members: (a) detection of the front surface of wales (highlighted in green) employing the RANSAC for plane detection; (b) detection of ties-braces (highlighted in gray) employing the DBSCAN; and (c) identification of the direction of the third principal axis comparing studs (highlighted in purple) and wales (highlighted in green) (side view).*

The second step is to identify the number of formwork system members in each category. Different categories of formwork system members have specific layouts. Formwork system members such as studs are installed along the second principal axis (Figure 6 (a)), and formwork system members such as wales are installed along the first principal axis (Figure 6 (b)). In other words, for a category of formwork system members that are installed along a specific principal axis, when creating a histogram of the number of points along this principal axis, a peak in the histogram represents a formwork system member, and it can be easily distinguished from lower peaks representing member spacing beside it. Considering such characteristics, the multiple peak detection algorithm is utilized to identify the number of studs and wales because they have been previously segmented into different sets of points with all formwork system members in their respective categories. For the multiple peak detection algorithm, the number of points for studs or wales is averaged into a baseline over the range of the second or first principal axis, and the number of points in each set bin is counted sequentially. If the number of points first exceeds and then falls below the baseline, it is considered a stud or wale (Figure 6 (c) and (d)). Although ties and braces have been previously segmented into different sets of points with a single formwork system member, it is unknown to which category a set of points belongs. Since the number of points for a tie is significantly smaller than that for a brace, the point number comparison algorithm is utilized to identify the number of ties and braces. For the point number comparison algorithm, the number of points for ties and braces is averaged into a baseline over the total number of ties and braces, and the number of points for each tie and brace is counted sequentially. If the number of points for a formwork system member is smaller than the baseline, it is considered a tie; otherwise, it is considered a brace.

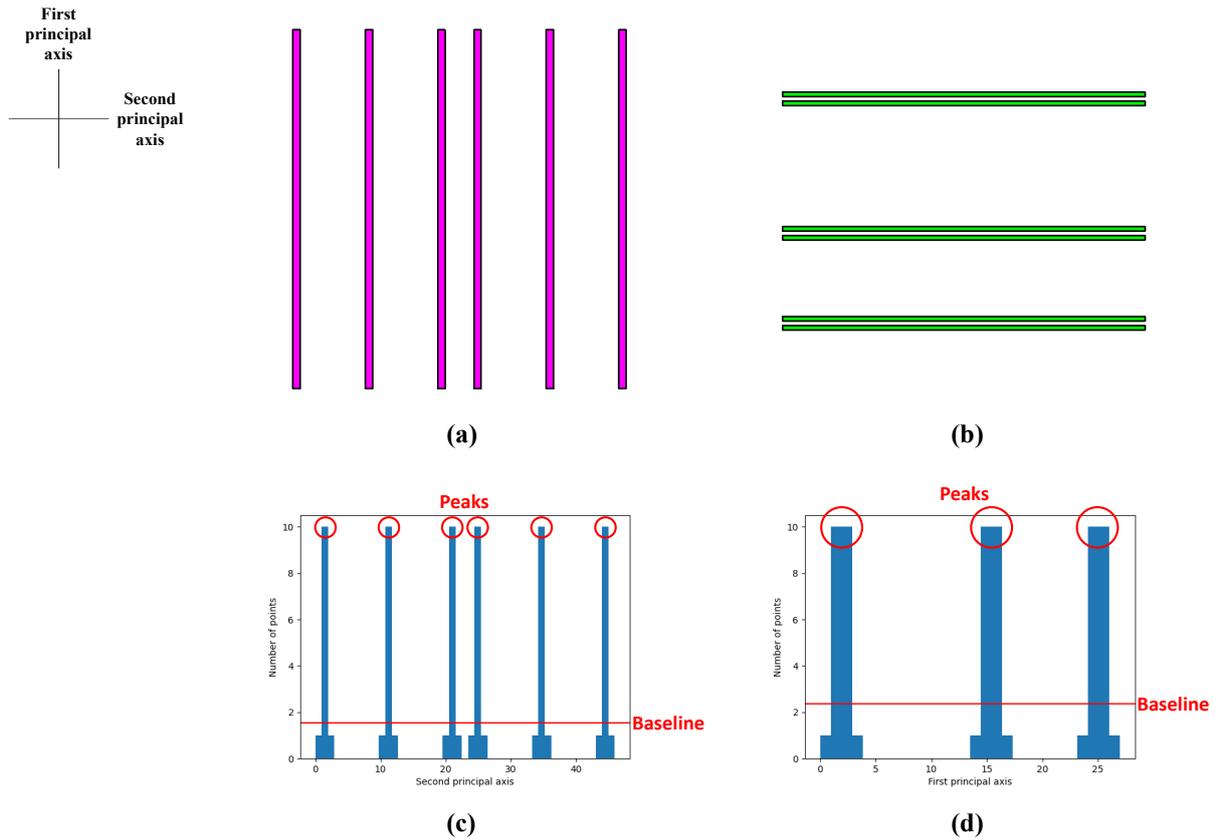

*Figure 6. Installation layouts of studs (a) and wales (b) from the front view; and identification of the number of studs (c) and wales (d) utilizing the multiple peak detection algorithm.*

The third step is to recognize formwork system members in each category. To facilitate the spacing measurement and the corresponding result positioning, each formwork system member in the same category needs to be detected and numbered. Although the positions of studs, braces, and wales can be approximately estimated with the histogram from the previous step, it is not accurate enough. Thus, the RANSAC for line detection is applied to detect each formwork system member, and a brace is required to be particularly detected twice as it consists of two poles. Since a category of formwork system members is installed along a specific principal axis, the relative position of each member can be determined according to the mean magnitude of the coordinates of its points on this principal axis when compared to those of other members in the same category. In this sense, formwork system members in each category are numbered sequentially in ascending order of mean magnitude for recognition. This process stops when the corresponding identified number is reached. For studs and braces installed along the second principal axis, the stud or brace with the smallest mean value on the second principal axis among its same category is located on the leftmost or rightmost from the front view and is numbered "1". The stud or brace with the second smallest mean value is located immediately to the right or left of it and is numbered "2", and the rest of the studs or braces can be deduced by analogy (Figure 7 (a) and (b)). For wales that are installed along the first principal axis, the wale with the smallest mean value on the first principal axis among its same category is located on the bottommost or topmost from the front view and is numbered "1". The brace with the second smallest mean value is located immediately to the top or bottom of it and is numbered "2", and the rest of the wales can be deduced by analogy (Figure 7 (c)). For ties that are not solely installed along the first or second principal axis, they are grouped depending on whether means are within the range of the wale numbered "1" on the first principal axis.

The tie with the smallest mean value on the second principal axis in the first group is located on the leftmost or rightmost from the front view and is numbered "1_1". The tie with the second smallest mean value in the first group is located immediately to the right or left of it and is numbered "1_2", and the rest of the ties in the first group can be deduced by analogy. Similarly, ties outside the first group are numbered sequentially as above (i.e., 2_1, 2_2, ..., 3_1, 3_2, ...) (Figure 7 (d)).

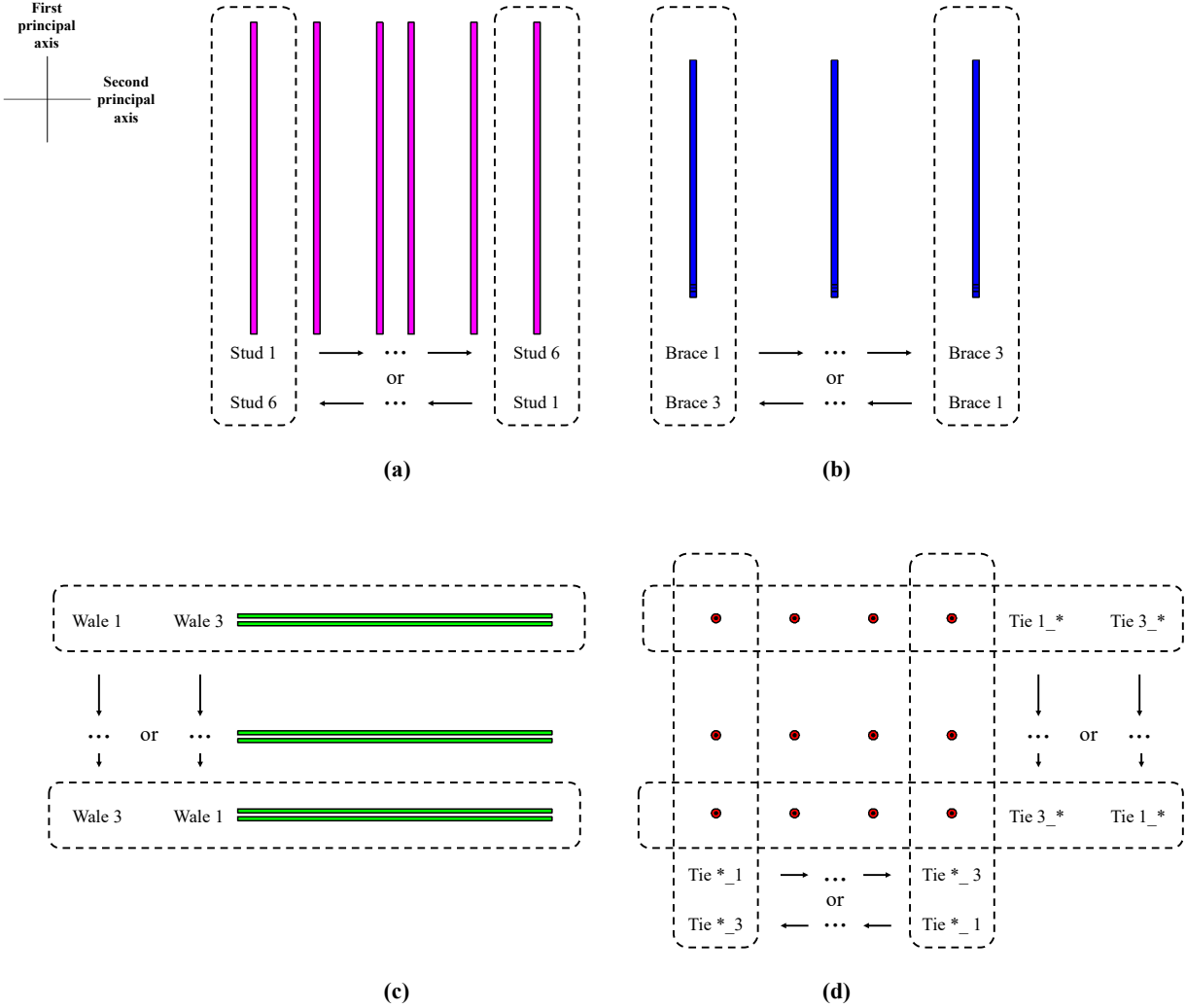

*Figure 7. Recognition of studs (a), braces (b), wales (c), and ties (d) from the front view.*

The fourth step is to measure the spacing between formwork system members in each category. For studs and braces that are installed along the second principal axis, the spacing to be measured corresponds to the horizontal spacing from the front view (Figure 8 (a) and (b)). For wales that are installed along the first principal axis, the spacing to be measured corresponds to the vertical spacing from the front view (Figure 8 (c)). For ties that are not solely installed along the first or second principal axis, the spacing to be measured corresponds to the horizontal spacing from the front view because their vertical spacing has been controlled by wales (Figure 8 (d)). In each category of formwork system members, spacing is determined by the difference between the means of the coordinates of points of two adjacently numbered members on the corresponding principal axis. As a result, the means of studs, braces, and ties are derived from the coordinates of points on the second principal axis, and those of wales are derived from the coordinates of points on the first principal axis.

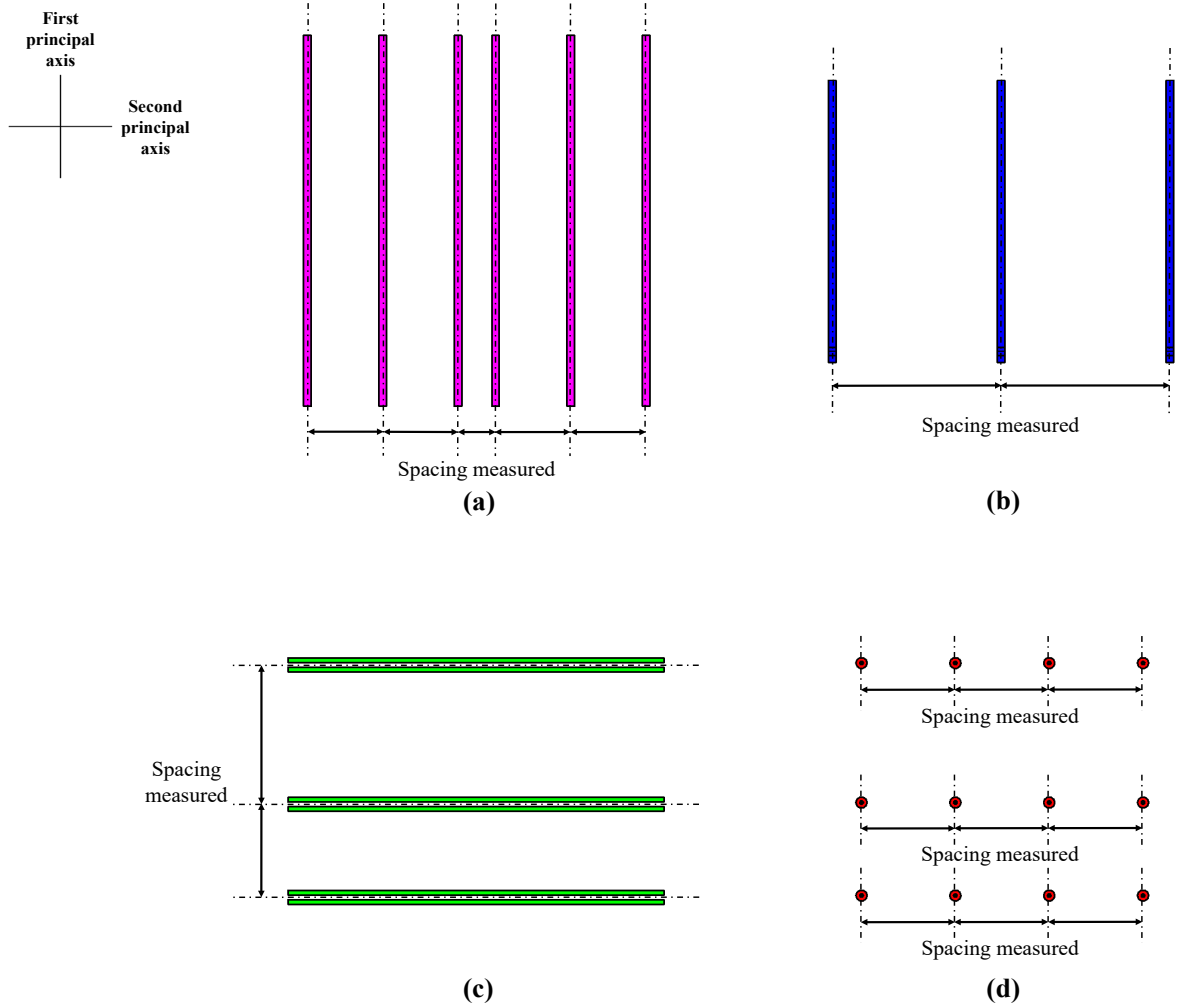

*Figure 8. Spacing measurement of studs (a), braces (b), wales (c), and ties (d) from the front view.*

## 4. Case study

To demonstrate the feasibility and effectiveness of the proposed framework, a formwork system on an ongoing construction project on a university campus was tested with a 3D laser scanner. The testing objectives were two sections of formwork systems. The first section included eleven studs, two wales, eight ties, and two braces, and the second section included twelve studs, two wales, and eight ties (Figure 9). The laser scanner used was a Leica BLK360 with a field of view of 360° (horizontal) and 300° (vertical), a scanning range of 0.6 m to 60 m, a point measurement rate of up to 360,000 pts/sec, and a 3D point accuracy of 6 mm@10m and 8 mm@20m [38].

In order to compare measurement results under different conditions, five cases were conducted with different testing objectives as well as numbers and distances of scans (Table 2). Case 1 with Testing objective 1 and Case 2 with Testing objective 2 adopted the same number of scans (i.e., 2), spacing of scans (i.e., 4.0 m), and distance of scans (i.e., 1.5 m). Cases 3, 4, and 5 with Testing objective 2 adopted the same number of scans (i.e., 1), but different distances of scans (i.e., 1.5 m, 3.0 m, and 4.5 m). Case 1 and Case 2, with different testing objectives, were used to verify the robustness of measurements. Case 2 and Case 3, with different numbers of scans, were used to verify the impact of the number on measurements. Case 3, Case 4, and Case 5, with different distances of scans, were used to verify the impact of the distance on measurements. Following the proposed framework, the five cases were processed with the help of the Open3D data processing library in Python [39]. Case 1 is used as an example to graphically display the details of the test, and the process for the other cases is similar. The results of all the tests are presented numerically.

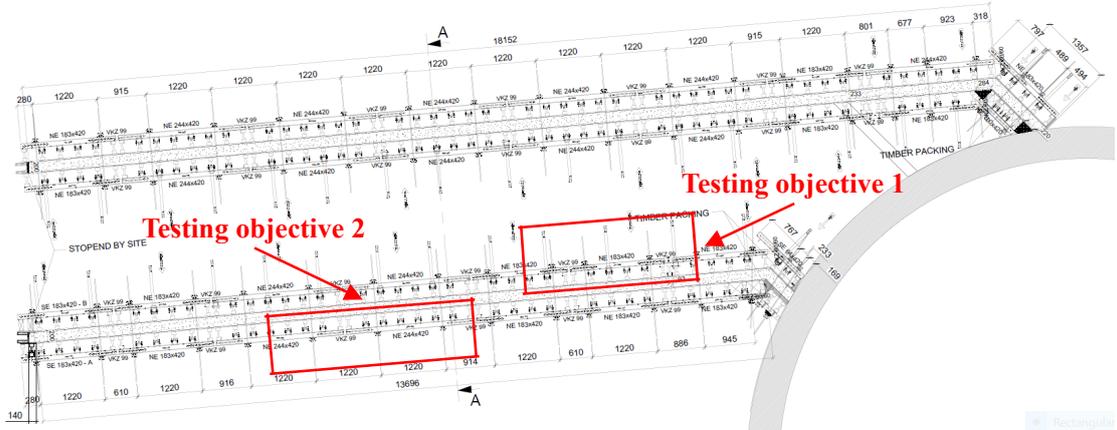

(a)

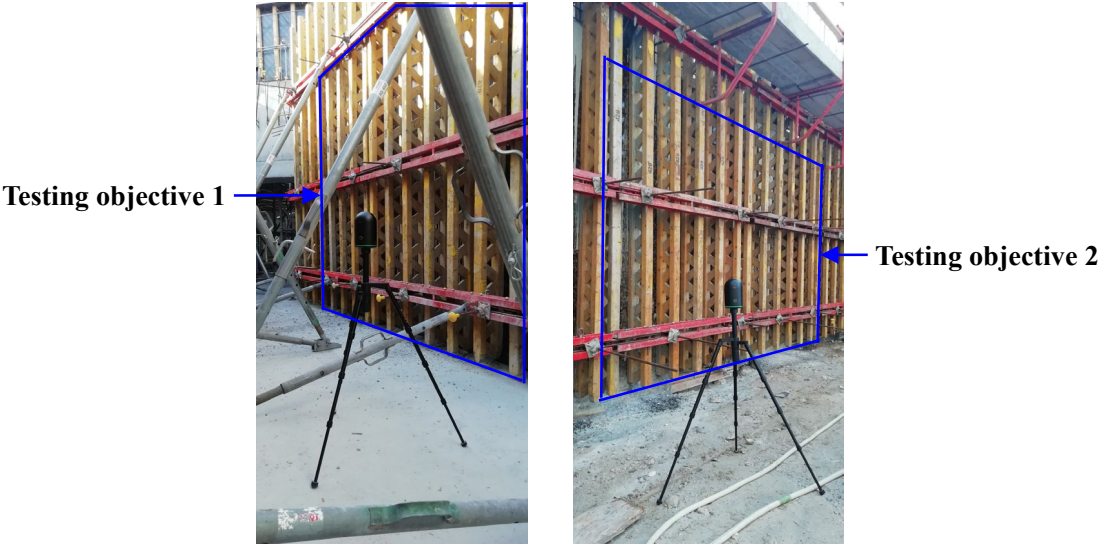

(b)

*Figure 9. Testing objectives 1 and 2 on the general arrangement drawing (a) and the construction site (b)*

*Table 2. Characteristics of the five cases.*

| Case | Testing objective | Number of scans | Spacing of scans (m)* | Distance of scans (m)** |
|---|---|---|---|---|
| 1 | 1 | 2 | 4.0 | 1.5 |
| 2 | 2 | 2 | 4.0 | 1.5 |
| 3 | 2 | 1 | - | 1.5 |
| 4 | 2 | 1 | - | 3.0 |
| 5 | 2 | 1 | - | 4.5 |

*Spacing in between locations of scanning positions
**Distance from the laser scanner to the testing objective measured perpendicularly

For Part 1, 3D point cloud data was preprocessed, as shown in Figure 10. First, the surroundings of the formwork system were removed from the raw point cloud (Figure 10 (a)). The thresholds of the pass-through filter were set according to the coordinates of points for the formwork system viewed on CloudCompare [33], and the raw point cloud was filtered to the desired portion (Figure 10 (b)). To remove the ground, the number of points in each bin with a size of 0.05 m was counted along the Z-axis of the histogram. The highest peak was considered as the ground level (Figure 10 (c)), and the points on the ground level and below were removed (Figure 10 (d)). Next, the outliers marked in red were removed from the point cloud (Figure 10 (e)). For the removal of the outliers, 100 neighbors were considered for computing the average distance for a given point, and the threshold based on the standard deviation of the average distances was set to 1. Then, the downsampling of the point cloud was performed by setting a voxel size of 0.01 m (Figure 10 (f)). Finally, the studs were segmented, and the coordinate system of the point cloud was transformed (Figure 10 (g)). The first, second, and third principal axes of the transformed coordinate system were the long side, short side, and normal directions of the detected stud surface, respectively. The RANSAC for plane detection was applied to segment the front surface of the studs. The distance threshold, the number of points that are randomly sampled to estimate a plane, and how often a random plane is sampled and verified were defined as 0.01 m, 3, and 1,000, respectively.

For Part 2, 3D point cloud data was processed and analyzed, as shown in Figure 11. First, the remaining categories of formwork system members were segmented based on the transformed coordinate system (Figure 11 (a)). The points marked in purple, green, and gray represent the segmented studs, wales, and ties-braces, respectively, and they were installed in turn along the positive direction of the third principal axis. In order to segment the wales and ties-braces, the distance threshold, the number of points that are randomly sampled to estimate a plane, and how often a random plane is sampled and verified were defined as 0.01 m, 3, and 1,000, respectively, when applying the RANSAC for plane detection; and the distance to neighbors in a cluster and the minimum number of points required to form a cluster were defined as 0.05 m and 30, respectively, when applying the DBSCAN. Next, the numbers of the formwork system members in each category were identified. The number of points for the studs was counted along the second principal axis, and eleven studs were identified according to the red baseline determined by the average number of points along this principal axis (Figure 11 (b)). The number of points for the wales was counted along the first principal axis, and two wales were identified according to the red baseline determined by the average number of points along this principal axis (Figure 11 (c)). The number of points for the ties and braces was counted, and eight ties and two braces were identified according to the red baseline determined by the average number of points over their total number (Figure 11 (d)). Then, the formwork system members in each category were recognized (Figure 11 (e)). The studs and braces were numbered sequentially according to the means of the coordinates of their points on the second principal axis from smallest to largest (i.e., Stud 1, …, and Stud 11; Brace 1 and Brace 2). The wales were numbered sequentially according to the means of the coordinates of their points on the first principal axis from smallest to largest (i.e., Wale 1 and Wale 2). The ties were numbered sequentially according to the means of the coordinates of their points on the X and Y-axes from smallest to largest (i.e., Tie 1_1, …, Tie 1_4, Tie 2_1, …, and Tie 2_4). Finally, the spacing between the formwork system members in each category was measured. For example (Figure 11 (f)), the spacing between Stud 1 and Stud 2 was the difference between the means of the coordinates of their points on the second principal axis.

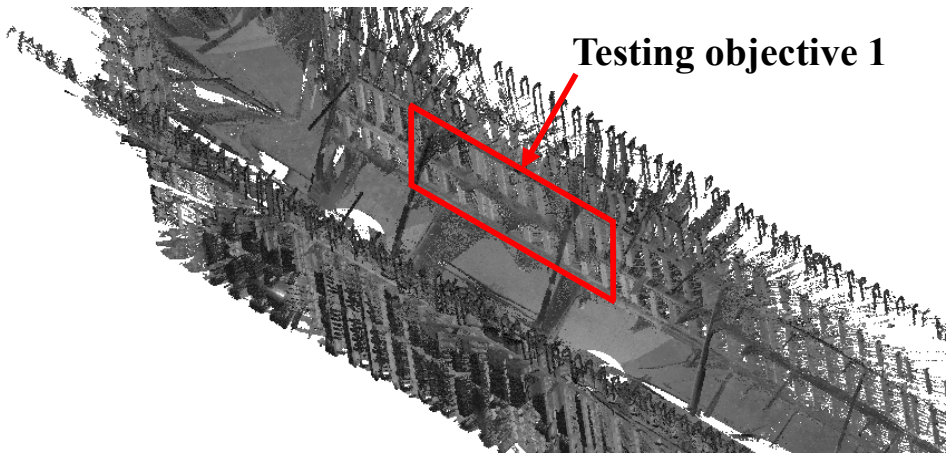

(a)

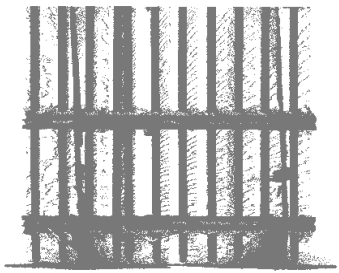

(b)

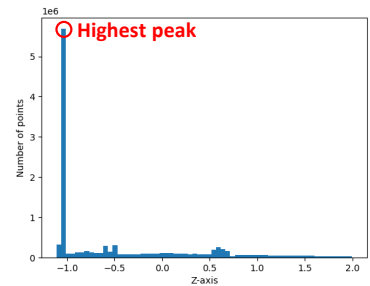

(c)

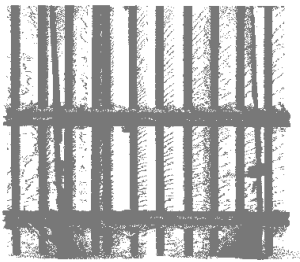

(d)

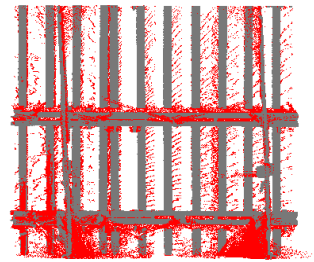

(e)

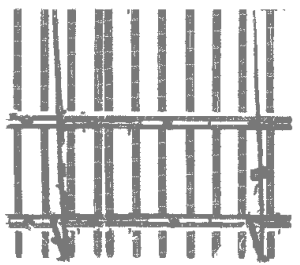

(f)

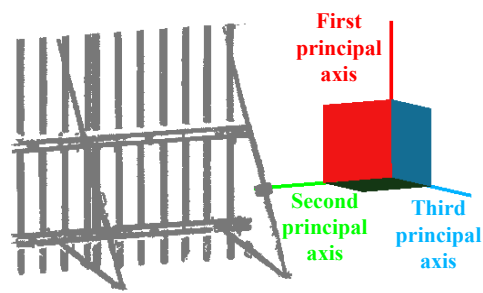

(g)

*Figure 10. 3D point cloud data preprocessing: (a) the raw point cloud; (b) the point cloud after removing the surroundings except for the ground; (c) the histogram of the number of points for the point cloud on the Z-axis; (d) the point cloud after removing the ground; (e) the point cloud after removing the outliers (highlighted in red); (f) the point cloud after performing downsampling; and (g) the point cloud after transforming the coordinate system.*

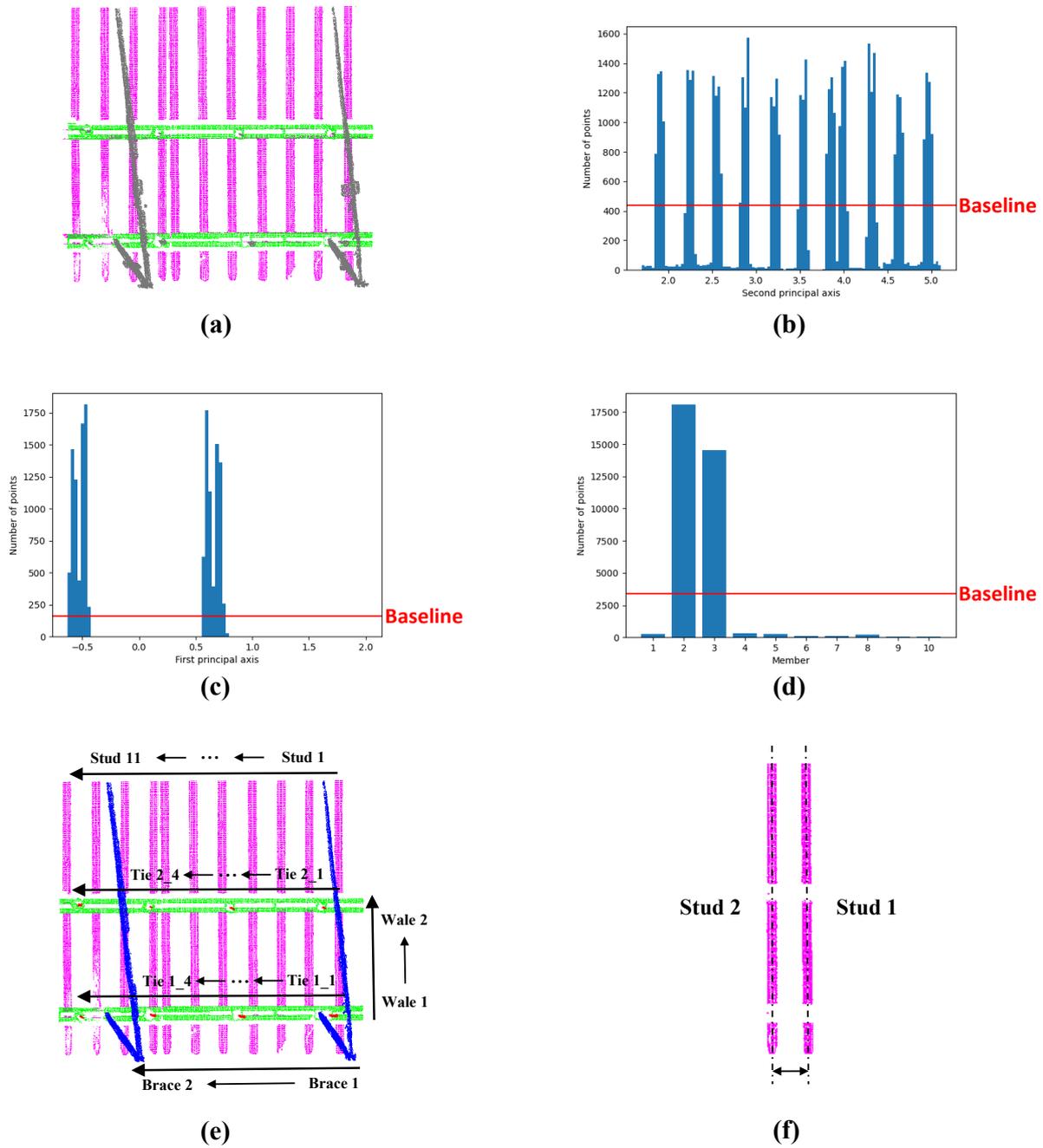

*Figure 11. 3D point cloud data processing and analysis: (a) the segmentation of the different categories of formwork system members; (b) the histogram of the number of points for the studs on the second principal axis; (c) the histogram of the number of points for the wales on the first principal axis; (d) the histogram of the number of points for the ties and braces; (e) the recognition of the formwork system members in each category; and (f) the spacing between Stud 1 and Stud 2.*

The comparison of the spacing measurement results using the 3D point cloud data ($R^{pc}$) and the traditional measuring tools (i.e., measuring tape and laser distance meter) ($R^{mt}$) is summarized in Table 3. The mean absolute error (MAE) and the mean absolute percentage error (MAPE) were used for the comparison metrics. The MAE was the mean of the absolute values of the differences between $R^{pc}$ and $R^{mt}$ (Equation 1). The MAPE was the mean of the percentages of the absolute values of the differences between $R^{pc}$ and $R^{mt}$ to corresponding $R^{mt}$ (Equation 2).

$$\text{MAE} = \frac{1}{n}\sum_{i=1}^{n}\left|R_i^{pc} - R_i^{mt}\right| \qquad (1)$$

$$\text{MAPE} = \frac{100\%}{n}\sum_{i=1}^{n}\left|\frac{R_i^{pc} - R_i^{mt}}{R_i^{mt}}\right| \qquad (2)$$

In each case, the MAE and the MAPE for each category of formwork system members were calculated. For example, in Case 1, the minimum (1.00 mm) and maximum (11.00 mm) MAEs were from the wale and the brace, respectively; and the minimum (0.08%) and maximum (0.52%) MAPEs were from the wale and the tie, respectively. Similarly, in each category of formwork system members, the MAE and the MAPE for each case were calculated. For example, in the stud, the minimum (1.60 mm) and maximum (2.45 mm) MAEs were from Case 1 and Case 3, respectively; and the minimum (0.45%) and maximum (0.78%) MAPEs were from Case 4 and Case 3, respectively. In addition, the MAE and the MAPE for all formwork system members in each case, as well as those for all cases in each category of formwork system members, were calculated. The minimum (2.39 mm) and maximum (4.17 mm) MAEs for all formwork system members were from Case 4 and Case 5, respectively; and the minimum (0.43%) and maximum (0.72%) MAPEs for all formwork system members were from Case 4 and Case 5, respectively. The minimum (1.80 mm) and maximum (11.00 mm) MAEs for all cases were from the wale and the brace, respectively; and the minimum (0.15%) and maximum (0.59%) MAPEs for all cases were from the wale and the stud, respectively. Overall, the MAE and MAPE for all formwork system members in all cases were 3.11 mm and 0.55%, respectively.

*Table 3. Comparison of the spacing measurement results using the 3D point cloud data and the measuring tools.*

| Formwork system member | Metric | Case | | | | | |
|---|---|---|---|---|---|---|---|
| | | 1 | 2 | 3 | 4 | 5 | All** |
| Stud | MAE (mm) | 1.60 | 1.64 | 2.45 | 1.36 | 2.09 | 1.83 |
| | MAPE (%) | 0.49 | 0.52 | 0.78 | 0.45 | 0.69 | 0.59 |
| Wale | MAE (mm) | 1.00 | 1.00 | 2.00 | 2.00 | 3.00 | 1.80 |
| | MAPE (%) | 0.08 | 0.08 | 0.17 | 0.17 | 0.25 | 0.15 |
| Tie | MAE (mm) | 4.83 | 5.33 | 4.17 | 4.33 | 8.17 | 5.37 |
| | MAPE (%) | 0.52 | 0.55 | 0.40 | 0.45 | 0.86 | 0.55 |
| Brace | MAE (mm) | 11.00 | - | - | - | - | 11.00 |
| | MAPE (%) | 0.46 | - | - | - | - | 0.46 |
| All* | MAE (mm) | 3.17 | 2.83 | 3.00 | 2.39 | 4.17 | 3.11 |
| | MAPE (%) | 0.48 | 0.51 | 0.62 | 0.43 | 0.72 | 0.55 |

*Results for all formwork systems
**Results for all cases

The comparison of the spacing measurement results with error bars is shown in Figure 12. The means are represented by blue dots, and the means ± standard deviation are represented by upper and lower red caps, respectively. The comparison of the stud, the wale, and the tie is shown in Figure 12 (a) and (b). The brace is removed from the comparison as there is only one result and, therefore, no variation. The performance of the wale was better whether in the MAE or the MAPE, and the performances of the tie and the stud were worse in the MAE and the MAPE, respectively. The reasons may be that: (1) the wale was large in size and less occluded by other categories of formwork system members, so the scanning effect was relatively good; (2) the tie was the smallest in size and made out of a higher reflective surface (i.e., metal), so the scanning presented less density and quality, and thus it performed relatively poorly in the MAE; and (3) the stud was not worst in the MAE but had a relatively poor performance in the MAPE due to its small spacing. The comparison of Case 1 and Case 2 with different testing objectives is shown in Figure 12 (c) and (d). The performance of the two cases was similar, indicating that the proposed framework is robust. The comparison of Case 2 and Case 3 with different numbers of scans is shown in Figure 12 (e) and (f). The performance of Case 2 was slightly better than that of Case 3, suggesting that the increase in the number of scans was beneficial in improving the scanning effect (as expected). The comparison of Case 3, Case 4, and Case 5 with different distances of scans is shown in Figure 12 (g) and (h). The performance of Case 4 was better than those of the other two cases, and the performance of Case 5 was worse than that of Case 3, which suggested that the distance of scans should be moderate, and too far was not good.

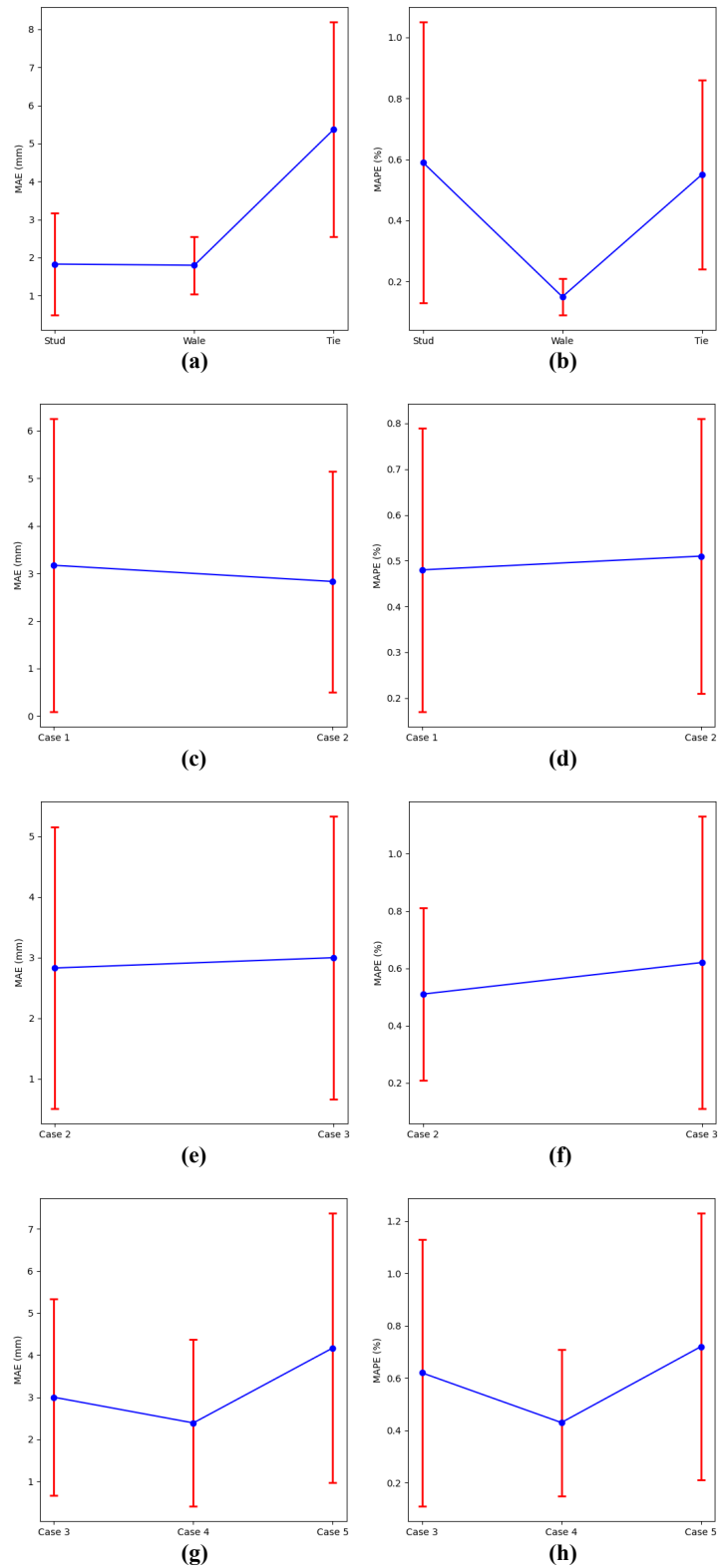

*Figure 12. Comparison of the spacing measurement results: the stud, the wale, and the tie ((a) and (b)); Case 1 and Case 2 with different testing objectives ((c) and (d)); Case 2 and Case 3 with different numbers of scans ((e) and (f)); and Case 3, Case 4, and Case 5 with different distances of scans ((g) and (h)).*

## 5. Discussion

The proposed framework for the spacing measurement of formwork system members with 3D point cloud data has been elaborated. Its feasibility and effectiveness have been demonstrated by a case study with different testing objectives and scanning numbers and distances. Considering the unique member and layout characteristics of the formwork system, the framework proposed the following three main contributions to enable a high degree of automation for the spacing measurement.

The first one is the determination of the orientation of the three principal axes of the transformed coordinate system. The orientation to fit formwork system members is a basis for realizing the automation of the spacing measurement, and it is used as a reference for the segmentation, identification, and recognition of formwork system members. Since the front surface of a stud is easier to be detected and has distinct shape features (i.e., length, width, and thickness are clearly different), it is used as the reference object for the transformation of the coordinate system such that the orientation of the three principal axes can be determined following the shape of the stud front surface. In this way, the transformed coordinate system adapts well to the orientation of formwork system members, making segmentation, identification, and recognition easier.

The second one is the determination of the number of formwork system members in each category. The known number of formwork system members is a prerequisite for the spacing measurement without human intervention. It is used to control the member detection, position identification, and spacing measurement processes, which can be stopped automatically when the corresponding number is reached. Studs and wales are installed along the second and first principal axes, respectively; therefore, the number of points for studs and wales varies regularly along the corresponding principal axes. Based on this fact, the numbers of studs and wales can be determined according to the variation situation on the baseline fixed by the average number of points over the range of the corresponding principal axis. The size of the ties is considerably smaller than that of the braces; hence, the number of points for ties and braces is significantly different. Taking advantage of this reality, the numbers of ties and braces can be determined according to the variation situation on the baseline fixed by the average number of points over the total number of ties and braces.

The third one is the determination of the relative position of formwork system members in each category. Only by determining the relative position of members in a formwork system can two adjacent members required for the spacing measurement be identified automatically. Therefore, the fact that a given category of formwork system members is installed along the corresponding principal axis (axes) is applied. Since studs are installed along the second principal axis, that implies that wales are installed along the first principal axis, braces are installed along the second axis, and ties are installed along the first and second principal axes. Their relative positions are determined based on the mean magnitude of the coordinates of points for each member on the corresponding principal axis (axes) when compared to those of other members in the same category. Accordingly, the two adjacent members required for the spacing measurement can be identified according to the labeled sequential number in ascending order of the mean magnitude.

## 6. Conclusion and outlook

With the progress and development of point cloud-related technologies and devices, 3D point cloud data is being applied to multiple dimensional quality inspections in construction projects, and has obtained positive acceptance because of the favorable representation manner of the 3D point cloud for the external surface of objects. However, the application of 3D point cloud data has rarely been studied to check the installation of formwork systems and, in particular, to measure the spacing of formwork system members including studs, wales, ties, and braces. Considering that the current spacing measurement is still mainly manual, and previous studies for other objects do not apply to the unique members and layouts of the formwork system, this research proposes a framework for the spacing measurement of formwork system members with 3D point cloud data to enhance its automation. In summary, the proposed framework is divided into two parts. The first part is 3D point cloud data preprocessing, consisting of four steps: (1) Remove surroundings, (2) Remove outliers, (3) Perform downsampling, and (4) Transform the coordinate system. The second part is 3D point cloud data processing and analysis, consisting of four steps: (1) Segment remaining members, (2) Identify the number of members, (3) Recognize members, and (4) Measure the spacing between members. A formwork system on an ongoing construction site was used to demonstrate the feasibility and effectiveness of the proposed framework. Five cases were considered with different testing objectives and scanning numbers and distances. When the 3D point cloud data approach was compared to the manual approach with a measuring tape and a laser distance meter for the five cases, the mean absolute error (MAE) and the mean absolute percentage error (MAPE) were 3.11 mm and 0.55%, respectively. Although the removal of surroundings around the formwork system required some manual operations, other steps, including the segmentation, identification, and recognition of formwork system members, were fully automated (i.e., without human intervention). The proposed framework was validated effectively, indicating that the 3D point cloud data approach is a promising solution and can potentially be an effective alternative to the manual approach.

There are some limitations to be further addressed in the proposed framework. On construction sites, it is inevitable to scan the surroundings of the formwork system. The points representing the surroundings are removed by applying the pass-through filter in the proposed framework. In this case, thresholds need to be determined manually by conforming to the 3D coordinates of the formwork system within the point cloud, which results in a prolonged step. In computer vision, machine learning techniques, including deep learning, have shown an excellent capability for object detection. It may be interesting to apply relevant techniques to detect formwork systems while excluding surroundings in the raw point cloud. In addition, some algorithms used in the proposed framework are required to set the corresponding computational parameters in advance. For example, the statistical outlier removal needs to set the number of neighbors and the threshold level based on the standard deviation of the average distances. However, the setting of such computational parameters relies largely on the scanning quality, and results may be different using the algorithm with the same computational parameters to process point clouds with different scanning quality. Therefore, an algorithm with the same computational parameters used for all point clouds without tuning is unreliable, especially on a construction site that easily suffers from surrounding factors (e.g., interference of different construction elements or activities). It will be valuable and challenging to develop more robust alternative algorithms. The above limitations could be investigated in future work.